# A Console GRID Leveraged Authentication and Key Agreement Mechanism for LTE/SAE

Rajakumar Arul, *Member, IEEE*, Gunasekaran Raja, Ali Kashif Bashir and Junaid Chaudry, *Senior Member, IEEE,* Amjad Ali

*Abstract*— Growing popularity of multimedia applications, pervasive connectivity, higher bandwidth, and euphoric technology penetration among bulk of the human race that happens to be cellular technology users, has fueled the adaptation to Long Term Evolution (LTE)/ System Architecture Evolution (SAE). The LTE fulfills the resource demands of the next generation applications for now. We identify security issues in authentication mechanism used in LTE that without countermeasures might give super user rights to unauthorized users. The LTE uses static LTE Key (LTE-K) to derive the entire key hierarchy, i.e., LTE follows Evolved Packet System – Authentication and Key Agreement (EPS-AKA) based authentication, which discloses user identity, location, and other Personally Identifiable Information (PII). To counter this, we propose a public key cryptosystem named "*International mobile subscriber identity Protected Console Grid based Authentication and Key Agreement (IPG-AKA) protocol*" to address the vulnerabilities related to weak key management. From the data obtained from threat modeling and simulation results, we claim that the IPG-AKA scheme not only improves security of authentication procedures, it also shows improvements in authentication loads and reduction in key generation time. The empirical results and qualitative analysis presented in this paper proves that IPG-AKA improves security in authentication procedure and performance in the LTE.

## I. Introduction

Mobile devices are pervasive and popular. Their extensive adaptation and versatility in usage have triggered an ever-growing increase in demand for network bandwidth. To meet these demands, network operators are carrying on extensive upgradation of their network infrastructure around the globe. The Long Term Evolution (LTE) standard is a 4G wireless communication technology that provides a wide coverage area and offers high-speed data and wide network bandwidth [1]. Due to network spectral efficiency and capability, LTE is expected to be the backbone technology for the next generation networks including emergency response systems and worldwide interoperable public safety systems. The LTE standard has evolved from 3rd Generation Partnership Project (3GPP) which helped other standards to grow, i.e., High-Speed Packet Access (HSPA), System Architecture Evolution (SAE), and LTE-Advanced (LTE-A), etc.

LTE technology can support both enterprise and ordinary consumers. Though with several advantages, LTE has vulnerabilities and security loopholes making it a sizable target for cybercriminals [6, 7]. These vulnerabilities and loopholes require intensive considerations from research and development community. The list of frequently used abbreviated terminologies in this manuscript is presented as Table 1.

The LTE-SAE network architecture constitutes of two prominent components:

1. Evolved Universal Terrestrial Radio Access Network (E-UTRAN): The first part comprises of one or more evolved NodeBs (eNB) which is accountable for effective radio transmission and reception from/to the mobile users respectively.

2. Evolved Packet Core (EPC): The EPC provides the subscriber a session keying material for the respective eNBs. It also holds long term keys that are used for authentication and establishment of security associations.

Rajakumar Arul and Gunasekaran Raja are with NGNLab, Department of Computer Technology, Anna University, Chennai, India. (e-mail: rajakumararul@ieee.org; dr.r.gunasekaran@ieee.org ).

Ali Kashif Bashir is with the Faculty of Science and Technology, University of the Faroe Islands, Faroe Islands. (e-mail: dr.alikashif.b@ieee.org).

Junaid Chaudry is with College of Security and Intelligence, Embry-Riddle Aeronautical University, Prescott AZ, USA and Security Research Institute, Edith Cowan University, Joondalup WA, Australia.

Amjad Ali is with UWB Wireless Communications Research Center in the Department of Information and Communication Engineering at Inha University, South Korea. (email: amjad.khu@gmail.com)

TABLE I
FREQUENTLY USED ABBREVIATIONS

| Acronym | Explanation |
|---|---|
| LTE | Long Term Evolution |
| IMSI | International Mobile Subscriber Identity |
| CK | Cipher Key (CK) |
| IK | Integrity Key (IK) |
| MME | Mobility Management Entity |
| HSS | Home Subscriber Server |
| UE | User Equipment |
| LTE – K | LTE Key |
| eNB | evolved NodeB |
| EPS-AKA | Evolved Packet System Authentication and Key Agreement |
| AV | Authentication Vector |
| C-GRID | Console GRID |
| IPG-AKA | IMSI Protected C-GRID based AKA |

Major elements of a typical EPC are Mobility Management Entity (MME), Policy and Charging Rules Function (PCRF), Serving Gateway, Packet Data Networks (PDN) gateway, and Home Subscriber Server (HSS).

The session keys provide cryptographic protection to the session between User Equipment (UE) and E-UTRAN. Hence, protection of all the session keys is important and critical for integrity of the communication within. The keys that are stored in eNBs, should be restricted only to their active operational domain.

The key generation and their verification in E-UTRAN are performed through Evolved Packet System Authentication and Key Agreement (EPS-AKA) [7, 8]. We analyze that EPS-AKA protocol for LTE has numerous security issues. For instance: When a UE enters the LTE's eNB communication area, a mutual authentication between UE and MME is initiated, where Authentication Vectors (AV) containing *[RANDom Number (RAND), AUThenticatioN Token (AUTN)]* is shared with UE in unencrypted format [1, 9]. As RAND is a key parameter in Non-Access Stratum (NAS)-encryption key generation, an intruder who gains access to the unencrypted authentication vector can compromise the entire LTE connection. During every UE access request to LTE network, the Access Security Management Entity (ASME) in MME performs access security and EPS-AKA based key distribution functionalities [9, 10]. Even though EPS-AKA is claimed to be safe and efficient, it suffers from identity attacks, high bandwidth utilization, and communication cost [11]. Another major problem is absence of an appropriate authentication mechanism to avoid Personally Identifiable Information (PII) theft through International Mobile Subscriber Identity (IMSI) exposure.

The IMSI is the piece of information that is primarily responsible for the identification of a subscriber on a cellular network and is unambiguously related to a given UE. It is conjointly used for getting different details of the mobile unit within the Home Location Register (HLR) directly or within the Visitor Location Register (VLR) as regionally derived. To prevent eavesdropping and other attacks on the radio interface, the IMSI is sent as rarely as possible. Unique identity of a subscriber is known through the IMSI that helps to fetch the user subscribed services and location [1-6, 12]. As shown in Fig. 1[3], first three digits of IMSI are the Mobile Country Code (MCC) followed by the Mobile Network Code (MNC) and Mobile Subscriber Identification Number (MSIN). The information of IMSI is also contained in the Universal Subscriber Identity Module (USIM) / IP Multimedia Services Identity Module (ISIM) card.

The IMSIs are typically used by the network operator to examine the subscribers whether to allow the subscriber to use another network operator service or not. The IMSIs are usually transmitted as plaintext on air as shown in Fig. 2 [6]. This leads to a range of security issues like location leak attacks, insider attack and lack of privacy preservation [46]. These attacks exploit the execution flaws in LTE Radio Resource Control (RRC) protocol. This reconnaissance may result into an active attack [7]. In IMSI based active attack the precise location of the mobile user is identified using *trilateration* as well as Global Positioning System (GPS) coordinates. Moreover, Man-in-the-Middle (MitM) and Denial of Service (DoS) attacks in LTE persist due to IMSI vulnerability [13]. In the LTE network, EPS-AKA performs key management operations. Once the source key (LTE-K) is compromised [11], the whole key hierarchy becomes vulnerable as LTE-K is the seed for subsequent key derivatives. This LTE-K based vulnerability is coined as a single key problem in [14, 15].

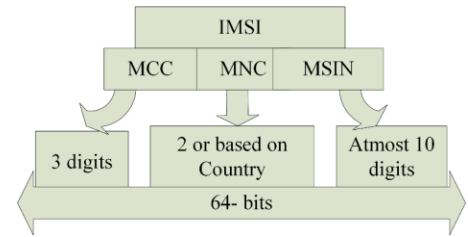

Fig. 1. Components of IMSI [3]

In this paper, we address the IMSI protection and static single key problem by proposing Identity (IMSI) Protected Console GRID (C-GRID) based Authentication & Key Agreement (IPG-AKA). It protects the IMSI using a public key based cryptosystem and solves the single static key problem that prevails in existing approaches like SE-AKA and EPS-AKA by utilizing a look-up table based approach. The IMSI protection in IPG-AKA is achieved by sending it as a cipher with a minimal number of network handshake messages and the proposed C-GRID acts as a seed for the LTE-K key hierarchy thereby solving the single key problem. The IPG-AKA outperforms EPS-AKA and compuationally less stranous to the middleware, because seamlessly imsersable and vividly scalable in the exisitng infrstrucutre. We used AVISPA tool with different attacker model such as chosen ciphertext attack, side channel attack, known plaintext attack and ciphertext only attack for the analysis.

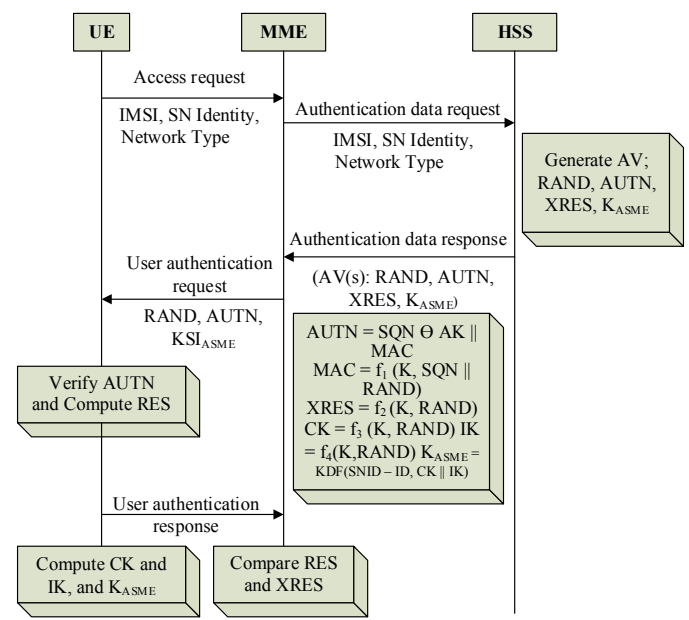

Fig. 2. EPS-AKA message exchange in LTE

This analysis prove that our protocol is secure against these attacks. Moreover, in comparison to EPS-AKA scheme, IPG-AKA improves security by 30%. Therefore, we can claim that IPG-AKA is secure against attacks and addresses the IMSI protection and static single key problems.

The rest of the paper is organized as follows: In section II, we present the literature review. In section III, we propose the LTE architecture that is equipped with the IPG-AKA. A discussion on privacy protection achieved through IMSI protected IPG-AKA is presented in section IV. In section V, we present the security analysis of IPG-AKA and performance analysis of the protocol in section VI. We present the concluding remarks in section VII.

## II. RELATED WORKS

Authentication is one of the most challenging security issues in advanced networks. The authentication process in LTE networks is primarily handled by EPS-AKA. Firstly, we will discuss the LTE authentication [1, 3, 5, 16] process using EPS-AKA. The LTE being a flat IP-based architecture is more prone to SQL injection, eavesdropping, IP address spoofing, MitM and PII theft [11], hence, the authentication process takes place at UE connection, call initiation, call termination, call handover, area updating etc. Also, numerous eNBs are managed by MME, and eNBs called low-cost base stations provide a direct path to malicious attackers. Once an adversary compromises the eNB, it can compromise the entire LTE access to the network [10, 14, 17].

A considerable volume of research is reported on the authentication [5, 35-41, 48] and on key agreement [18, 42-46] process to overcome the security gap that needs improvement in the recently proposed symmetric [13], and asymmetric security solutions [17]. Machine learning based authentication [2, 4, 46] and aggregation [47] is proposed to overcome the security flaws in the dynamic environments. Various security flaws in EPS-AKA are addressed in the literature. Firstly, *false base station attack* which is a redirection attack where a UE's attack request is redirected from the intended network to other networks separated by a long distance [19]. Secondly, the *impersonation attack* where an intruder is given a corrupted AV to impersonate any legitimate user [11, 19, 20]. Thirdly, usage *of synchronization approach* among UE and HSS where if periodic synchronization fails, it leads to synchronization error and communication failure [19, 20].

A derived AP-AKA mechanism to address these issues was proposed in which the UEs had the flexibility to select the authentication execution flow [20]. But the UE's serving network authentication was completely dependent on the home network server for performing the AV distribution when AVs were exhausted or key materials expired. In home networks, this approach certainly increases the computation load at the server and incurs access delay for mobile users. Another approach named X-AKA, an extension of Universal Mobile Telecommunications System Authentication and Key Agreement (UMTS-AKA) protocol was proposed to reduce the AV transmission overhead to improve bandwidth utilization in the standard EPS-AKA mechanism [21]. In this approach, a symmetric key based mechanism focuses on minimal bandwidth consumption and communication overhead to generate AV at the serving network instead of the home network [22]. However, these mechanisms do not prevent redirection and MitM attacks. For an environment with a large number of users, a secure group key approach incurs authentication delay due to numerous re-key construction requests and the extensive pre-processing steps involved in it [23].

An authentication and key agreement protocol were proposed for a group of mobile users who moves from a home network to the serving network. The reported results show near optimal authentication performance by compromising security [24, 25]. Message integrity protection against signal cancellation has been proposed using Manchester coding [3]. An execution environment adaptive protocol is proposed in [26] to reuse the available authentication vectors but the protocol is vulnerable to MitM, DoS, redirection, and PII compromise attacks. Introduction of public key based approaches will possibly be a way towards improving LTE security [28, 29]. The IMSI catcher is a hardware equipment that acquires all the IMSIs accessible close to it and are sent over the radio without encryption. The IMSI catching is avoided by securing IMSIs [30] but still location tracking and LTE-K vulnerability persists. In the LTE network, LTE-K is the source key from which the entire key hierarchy is derived to secure the full LTE communication [27]. If this LTE-K is compromised [11] at any level of communication, the whole key hierarchy becomes vulnerable. Our analysis of the key hierarchy reveals that the existing LTE networks are prone to compromise due to weak key management. This LTE-K based vulnerability is coined as a single key problem [14, 15]. So far, there are no sufficient solutions towards LTE security vulnerabilities.

Thus, major security issues like IMSI exposure and LTE-K attack poses active thread to PII. Public key based cryptosystem can address this issue; however, they only resolve half of the problem and single key problem still holds. With IPG-AKA, we aim to resolve this issue using a table based approach and by making key dynamic. Even if the adversary found the LTE-K derived through IPG-AKA, it cannot be reused to hack the entire network. The randomness of bit sequence generated through C-GRID makes it difficult to reuse in the entire network for the hacker, thereby making IPG-AKA a robust solution for handling single key problem of PII.

## III. THE LTE ARCHITECTURE WITH IPG-AKA

The LTE architecture comprises of primarily two components: EPC and EUTRAN. To ensure the integrity of the master key, we incorporate additional resources in IPG-AKA based authentication system in existing LTE architecture. The changes are kept to the least to minimize the effects on the existing architecture as shown in Fig. 3. The EPC framework includes LTE radio base stations (eNB) [31, 47] that connect to the control plane components such as MME, User Plane Gateways such as Packet Data Network Gateway (PDN-GW) & Serving GW (SGW). The MME acts as a Key Distributor (KD) [3-9] in the LTE. The EPS network holds both MME and SAE [1, 2] and connects these two to the E-UTRAN entities over the many-to-many interface via HSS.

A mutual handshake between the UE and the EPC is the most important security feature in the LTE security framework. Secure communication between EPC and UE takes place by sending a mutual authentication parameter of EPS-AKA over the network. Consequently, the UE communicates with the EPC through E-UTRAN networks. The UE communicates with the Public Switched Telephone Network (PSTN) through the 3GPP or E-UTRAN access networks via the EPC. The LTE supports the communication between the EPC and Non-3GPP network. However, there are two types of Non-3GPP access networks, namely trusted Non-3GPP access networks, and untrusted Non-3GPP access networks.

Based on policy enforced by the LTE network operator, a network is defined to be trustworthy or untrustworthy. Usually, for the Non-3GPP access networks, the UE communicates with the Authentication Authorization & Accounting (AAA) server by sending an untrusted Non-3GPP authentication *SWa* interface and trusted Non-3GPP authentication *STa* over the networks. The replication of the AAA server (proxy AAA server) is used for authenticating signaling during communication. So, for a trusted Non-3GPP the UE and the AAA server together send an EPS-AKA signaling, whereas, in untrusted Non-3GPP, the UE sends through an evolved Packet Data Gateway (ePDG) for confidence over the network to the EPC. The UE and the ePDG perform IP security tunnel establishment known as Internet Key Exchange v2 (IKEv2) with EPS-AKA. During tunneling, the UE and the AAA server receives communication from EPC mutual authentication of the user and the EPC.

The LTE is fully optimized for packet data access and it also provides Quality of Service (QoS) in voice communications. The comparison between the LTE radio and UMTS functions is best viewed in LTE architecture explicitly in terms of higher user data plane bandwidth, longer UE connected state duration, use of Discontinuous Reception in the connected state, Medium Access Control (MAC) level identity and Radio Network Controller (RNC). Even though the 3GPP committee has framed the stringent mobility procedures, vulnerabilities still exist during LTE handover [32] and seamless connectivity procedure. There is certain handover [33] procedures as S1 and X2 wherein S1, the message exchange occurs multiple times by contacting MME. It brings out a longer handover delay which in turn results in performance degradation as the location between the MME, and Home eNB (HeNB) is far away. In X2, the handover is done among eNBs.

A new key management mechanism is devised for the handover keying purpose. There are two different ways for the keying process, vertical and horizontal keys to derive the new eNB keys. To achieve secure communication between a user and eNB, an MME and the UE shall derive a $K_{eNB}$ and a Next Hop (NH) parameter from the $K_{ASME}$, which gets derived by the UE and the MME, whereas NH Chaining Counter (NCC) is associated with each $K_{eNB}$ and the NH parameter. The $K_{ASME}$ derives the $K_{eNB}$, and it gets associated with a virtual NH parameter with NCC value that is equal to zero. During the initial set-up, both the UE and the eNB use $K_{eNB}$, which is utilized for the secure communication in on-air surface, whereas during handover session keys are employed. Here in this set-up UE and eNB with the use of session keys produce the $K_{eNB}$ in LTE for the secure communication.

All the above-mentioned keys are derived from the static LTE-K that stays constant throughout the LTE communication. Thus, the impact of a compromised LTE-K may compromise all derived keys [11]. In other words, the single key problem in LTE can lead to the compromise of the entire key sets derived during an LTE lifetime for a UE. To address the static LTE-K problem, we propose a table based dynamic key generation mechanism (IPG-AKA), which in turn changes LTE-K at a periodic interval of time. A 5x5 or 7x7 C-GRID must be maintained for every UE in both UE and its corresponding MME as shown in Fig 3.

This C-GRID acts as the source material to generate a different set of LTE-K to enhance LTE security. The LTE key hierarchy and its security are enhanced through a C-GRID based key generation as shown in Fig. 4. The proposed IPG-AKA grid hierarchy includes various keys such as

- $K_{eNB}$ - a security key generated from $K_{ASME}$ by UE and MME
- $K_{NASint}$ - a NAS traffic integrity protection key derived by UE and MME from $K_{ASME}$
- $K_{NASenc}$ - a NAS traffic encryption key derived by UE and MME from $K_{ASME}$
- $K_{UPenc}$ - a User Plane traffic protection based encryption key derived from $K_{eNB}$ by UE and eNB
- **$K_{UPint}$** - a User Plane traffic integrity protection key derived from $K_{eNB}$ by Donor eNB (DeNB) and Relay Nodes (RN)
- $K_{RRCin}$ - an RRC traffic protection integrity key using Key Derivation Function derived by UE and eNB from $K_{eNB}$
- $K_{RRCenc}$ - an RRC traffic encryption key derived by UE and eNB from $K_{eNB}$

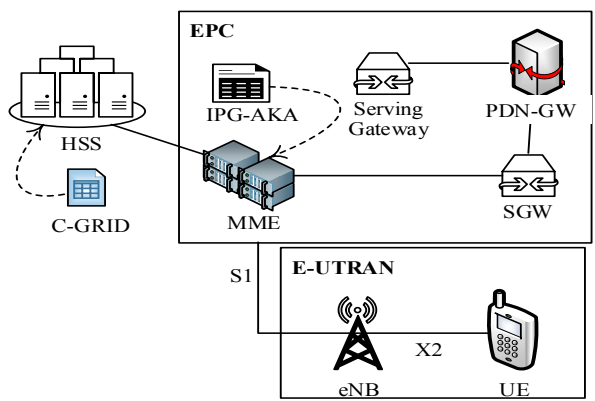

Fig. 3. LTE Architecture with IPG-AKA

Here, NH ensures forward key security in the LTE that is agreed upon by UE and MME. Finally, the key $K_{eNB}^*$ supports horizontal and vertical key derivation in LTE that was derived by UE and eNB. A sample C-GRID from which the LTE-K is generated is shown in Table II.

The C-GRID is populated with a set of binaries that are multiples of 8. The grid can be any '*n x n*' matrix; here a *5 x 5* and a *7 x 7* C-GRID are considered to brief the function of the key grid and the following LTE-K derivation. Each position in the C-GRID holds binaries denoted by a symbol as reflected in Table II. The '*ϕ*' value in the key grid always accounts for a null value. Any combination of the characters of the alphabet fetched from the grid will be fed as input to the proposed C-GRID based key generation function to generate a 256 bits dynamic LTE-K before every authentication initiation.

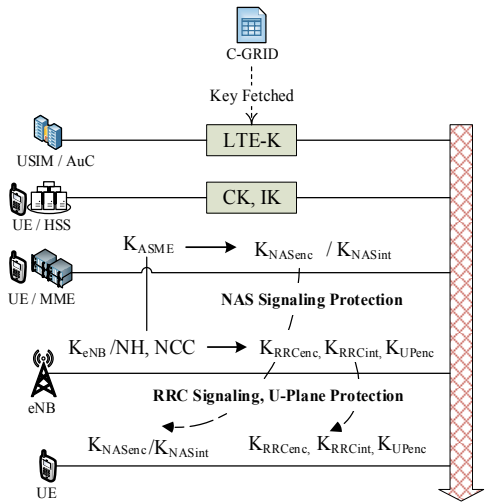

Fig. 4 Pseudo dynamic LTE-Key Hierarchy

TABLE II
C-GRID WITH ACTUAL SEEDS

| 8 bits | 16 bits | 24 bits | 16 bits | 8 bits |
|---|---|---|---|---|
| ϕ | B | E | H | T |
| A | M | ϕ | C | D |
| Q | F | J | G | ϕ |
| V | ϕ | S | L | K |
| R | W | B | ϕ | S |

TABLE III
SEED POSITIONS IN C-GRID

| 8 bits | 16 bits | 24 bits | 16 bits | 8 bits | |
|---|---|---|---|---|---|
| $a_{11}$ | $a_{12}$ | $a_{13}$ | $a_{14}$ | $a_{15}$ | |
| $a_{21}$ | $a_{22}$ | $a_{23}$ | $a_{24}$ | $a_{25}$ | |
| $a_{31}$ | $a_{32}$ | $a_{33}$ | $a_{34}$ | $a_{35}$ | |
| $a_{41}$ | $a_{42}$ | $a_{43}$ | $a_{44}$ | $a_{45}$ | |
| $a_{51}$ | $a_{52}$ | $a_{53}$ | $a_{54}$ | $a_{55}$ | |

**Algorithm 1: C-GRID Generation**

**Input:** Variable bit elements
**Output:** C-GRID as a table
1: **Start** GRID Generation
2: **Initiate** '*n x n*' matrix as GRID Layout
3: Setup GRID as variable bit columns
4: **for** $i^{th}$ column ($2^3*i$) sets the capacity of the column
5:    **for** each column **do**
6:       null($\phi$) value placement
7:    **end for**
8: **end for**
9: **Stop** GRID formation
10: **Stop** GRID Layout
11: Flood GRID with binaries
12: **Stop** GRID Generation

The placement of '*ϕ*' in the key C-GRID uniquely differentiates the key grid among different users. The binaries required for LTE-K generation are fetched using the alphabet position value in the C-GRID as given in Table III. A sample LTE-K generation sequence from a *5 x 5* key grid is, *24*4+16*4+16*4+8*4 = 256* bits. The combination of the bits can be fetched from any column as per the designed dynamic key feeder function.

*A. IPG-AKA C-GRID generation algorithm*

*Step 1*: Initiate a *5 x 5* or *7 x 7* matrix based grid at the UE and HSS end.

*Step 2:* Place a null value '*ϕ*' randomly such that each row and column in C-GRID must have not more than one (*ϕ*) value.

*Step 3:* Populate rest of the C-GRID positions with random alphabets.

Step 4: Assign binary values to every alphabet on the predefined binary size as *8 bits, 16 bits, 32 bits (*$2^3$* and above)* respectively.

*Step 5:* Placement of mirror columns with same bit size in the grid helps us to increase the complexity of the grid and every grid must possess at least (*n/2 – 1*) mirror column.

*B. IPG-AKA C-GRID based key generation function*

Once a C-GRID is generated, a secret Key Sequence (KS) is formed for the C-GRID using a secret Key Feeder Function (KFF). Both KS and KFF are pre-shared along with the C-GRID in UE and HSS. The secret key sequence and feeder vary for every UE and an eNB pair. The secret key generation process based on a *5 x 5* C-GRID is described for a sample secret key feeder, "*x = ((x*y+i)*(1024+i)*(i*j)) mod 4*".

Assume that the secret key sequence KS, which is going to replace the static LTE key, takes the form: [$C_3, C_1, C_4, C_3, C_2, C_4, C_1, ....C_2, C_4, C_3$] whose size is *256* bits. The steps involved in dynamic LTE key generation from Table IV are as follows:

*Step 1:* Find the null value '*ϕ*' from columns $C_1$ to $C_5$ in C-GRID as shown in Table IV.

TABLE IV
C-GRID BASED LTE-K GENERATION

| 8 bits | 16 bits | 24 bits | 32 bits | 24 bits | 16 bits | 8 bits |
|---|---|---|---|---|---|---|
| F | C | A | W | φ | R | W |
| G | E | L | O | K | φ | M |
| A | φ | E | P | S | G | S |
| U | H | B | S | L | Q | φ |
| R | K | N | φ | E | Z | H |
| φ | Q | K | D | B | Y | K |
| M | D | φ | J | N | H | Q |

---

**Algorithm 2: Key Sequence Formation**
**Input:** Elements from C-GRID
**Output:** Key Sequence

1: **Start** *Key_Sequence frame*
2: **Initiate rand()** for the $2^x$ bits value
3: Form 256_bits pattern in multiples of $2^x$ fields
4: Mix iterations like $2^2*4+2^3+2^4*2...$
5: **Store** Sequence
6: **Stop** *Key_Sequence frame*

---

*Step 2:* Compute the secret KFF for every column iteratively excluding the fields having a null value to fetch random binaries from C-GRID.

*Step 3:* For every successive iteration, the KFF returns positions in C-GRID, which are mapped randomly to the pre-defined KS for a successful LTE Key generation. Binaries are to be taken from returned KFF values to generate the LTE Key. The sample procedure of a key sequence fetched from Table IV is presented in Algorithm 4.

### C. Properties of the Key generated from IPG-AKA C-GRID

*Repeatability:* The sequence of numbers generated from the C-GRID possesses a unique generation technique, which produces different values for the same seed or initial value.

*Randomness:* All the 256 bits generated through IPG-AKA scheme is independent and are uniformly distributed random variables that pass all the statistical tests for randomness.

*Storage efficiency:* Key sequence generation only follows look-up operation and the space required for storing the resultant values retains the same; so, IPG-AKA is more efficient in terms of storage & key generation when compared with the existing approaches

*Disjoint sub sequences:* In the proposed IPG-AKA, by having a single key or a part of the key, next key cannot be generated and no correlation exists between simulations with different initial seeds.

*Long Period:* If a pseudo-random number sequence uses finite precision arithmetic, the sequence will repeat itself with a finite period and this will be much longer than the number of random numbers needed for the simulation.

*Insensitive to seeds:* C-GRID formed in IPG-AKA scheme is insensitive to the initial seed value since, for the same seed IPG-AKA generated two different values, thereby indicating non-dependency of the period and randomness.

*Portability:* In IPG-AKA, by having the same C-GRID and key generation function we can generate the same key sequence in 'n' number of machines. This conveys the evidence of generating the same key if C-GRID & key generation function matches.

*Homogeneity:* All the keys generated from IPG-AKA scheme imply the randomness of the sequence of all bits.

### IV. THE PROTECTED IMSI DATA

The first dispute resolved through IPG-AKA is single static key problem and the second issue addressed in this paper is the protection of PII. User privacy is the key problem that needs immediate attention in the future interconnected networks. In LTE network, the PII of users is vulnerable through the exposure of IMSI over the air. When an adversary gains access to an IMSI, it is possible to track all the services subscribed by that user. In the proposed scheme shown in Fig. 5, the IMSI is sent over the air implicitly by concealed it with nonce through following steps:

*Step 1:* When a UE attempts to enter the LTE network's range, the MME sends a user identity request to the UE with the following four parameters as a large prime *'p'*, two

---

**Algorithm 3: Key Generation from C-GRID at HSS**
**Input:** C-GRID
**Output:** LTE-K

1: **Start** *Key_Generation_256 bits*
2: **Initiate** *Key_Sequence Fetch*
3: **if** (Key_Sequence = Found)
4:     Fetch Key_Sequence
5: **else**
6:     **if** (New Member = 'Yes')
7:         Frame new *Key_sequence*
8:     **end if**
9: **else**
10:    Reject
11: **end if**
12: **Finish** *Key_Sequence Fetch*
13: **Start** *Key_Sequence Splitting*
14: **Divide** *Key_Sequence* into *Alpha_fields*
15: **Initiate** Bitwise Fetch on Alpha
16: **for** Each *Alpha_fields* **do**
17:    **if** (Previous value Fetched is from $i^{th}$ column) **then**
18:        fetch Next Value from $n-i^{th}$ column
19:    **end if**
20: **Finish** Alpha_fields Fetch
21: **Stop** *Key_Sequence Splitting*
22: **Merge** *Key_Sequence* by *Alpha_fields*
23: **Stop** *Key_Generation_256 bits*

---

**Algorithm 4: Key formation sample**
**Input:** C-GRID
**Output:** LTE-K

1: **for** (i = 0 to n)
2:    **for** (j = 0 to n-1)
3:        **While** (C[j][i] != '*φ*')
4:            y = (y * x) + (y + x)
5:            x = ((x * y + i) * (1024 + i) * (i * j)) mod 4
6:            select n variables
7:        **end while**
8:        **if** (required variable selected == 'Yes')
9:            Exit
10:       **else**
11:           **go to** step 3
12:       **end if**
13:   **end for**
14: **end for**

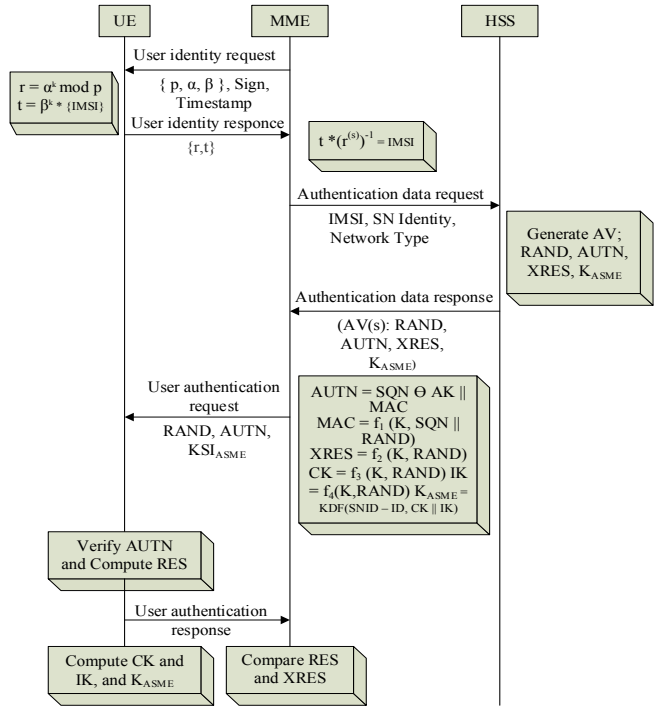

Fig. 5. Message exchanges of IPG-AKA

computed variables *(α, β)*, a time stamp and its signature signed by a trusted authority to support the MME authentication to UE.

*Step 2:* The UE computes *[r = α^k mod p and t = β^k * IMSI]* and sends the computed values *(r,t)* in the user identity response message. Here the computations are done with Mobile Equipment's (ME) IMSI and therefore the IMSI is shared with the MME in a secured way.

*Step 3:* On receiving the user identity response the MME computes the inverse process as $t * (r^s)^{-1} = IMSI$ where *'s'* is the secret key pair of *'β'* and hence the IMSI is obtained at the MME. Followed by this, the MME forwards the BS authentication data request to HSS with the IMSI, SNIdentity and the UE's network type.

*Step 4:* The HSS performs the SNID authentication, and on success, it retrieves the LTE-K from its database and computes the $K_{ASME}$ using sequence number, RAND, AUTN and gains an authentication vector. Then the authentication vector with RAND, AUTN, $K_{ASME}$, eXpected RESponse (XRES) is sent to MME in the authentication data response message.

*Step 5:* The MME fetches the RAND, AUTN from the received AV and adds $KSI_{ASME}$ and forwards these parameters in the user authentication request message to UE.

*Step 6:* The UE computes the Cipher Key (CK) and Integrity Key (IK) using its LTE-K and received RAND. On receiving the $KSI_{ASME}$ value, the UE can generate the similar $K_{ASME}$ key at the UE side. Finally, UE sends the RESponse (RES) message to MME.

*Step 7:* The MME on receiving the UE response message, compares the RES parameter with its XRES value; if it matches, then the UE is authenticated successfully or else the access request will be denied by the MME.

## V. SECURITY ANALYSIS

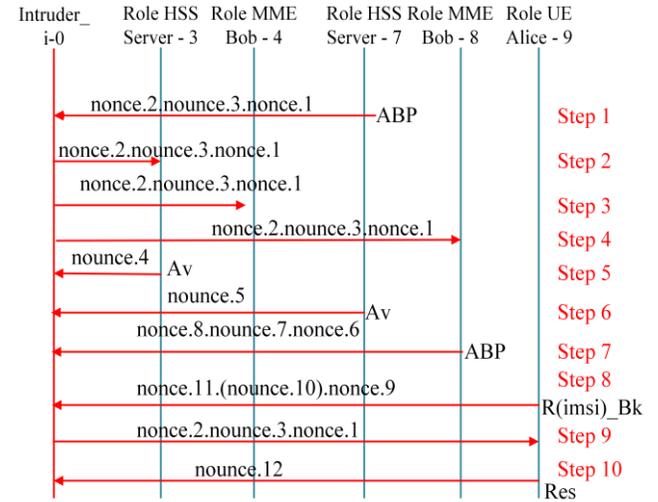

Fig. 6. Intruder simulation for IPG-AKA

The security verification of IPG-AKA mechanism is done using AVISPA [34] tool. The entities involved in the proposed IPG-AKA mechanism are ME/UE, MME, and HSS. The IPG-AKA protocol specification comprehends an environment that includes the transition to every role, corresponding sessions, the intruder knowledge, and the environment. The protocol is verified for a specific secrecy and authentication goal specified in goal section. Finally, the IPG-AKA protocol is checked with all back-end servers using various attacker models like known cipher text attack and cipher text only attack for key secrecy & authentication. The results have shown that the proposed IPG-AKA mechanism is safe and has bounded some sessions. The intruder simulation scenario in which the proposed IPG-AKA verification was done is shown in Fig. 6.

### A. Computational Complexity of C-GRID

The computational complexity of the C-GRID based LTE-K generation is analyzed with a *5 x 5* sample grid consisting of 360 bits *(8*5+16*5+24*5+16*5+8*5 = 360 bits)*. This actual grid size includes the NULL value in every grid column. In a practical scenario, intruder requires $2^{256}$ iterations to identify a security key of 256 bits size that is derived from the grid.

However, with the IPG-AKA mechanism, the complexity of any intruder guessing the security key is increased from $2^{256}$ to $2^{360}$ iterations for a *5 x 5* grid, as he must break both the grid and key to generate future keys. The intruder key computation complexity increases with increase in the grid size. The LTE-K secret key derived from the grid, being valid for a short time span is unbreakable in polynomial time, i.e. time taken to compute $2^{256}$ iterations. Added to this advantage, the complexity of the C-GRID is further enhanced by NULL value placement in every grid column where the null value placement position varies for every column of a grid and varies across different grids. To locate the NULL value's position in a single grid column, it takes "$2^{bitsize}$ * *number of rows * number of characters of the alphabet*" number of computations. Additionally, a *5 x 5* GRID contains $2^{288}$ values excluding the null values, which creates $2^5$ times more than the required $2^{256}$ bits key. Therefore, the illegitimate calculation of security key from the C-GRID in IPG-AKA mechanism is not possible.

### B. Mathematical Analysis of C-GRID

The security breach in IPG-AKA mechanism is evaluated and the breach time is computed as shown in Equation (1).

$$Breach\ Time = \Sigma(\mu_B * E_C * E_R * E * N_V) \qquad (1)$$

where,

$\mu_B$ = Number of bits ∈ Exponent
$E_C$ = Number of elements ∈ a column
$E_R$ = Number of elements ∈ a row
$E$ = All possible elements ∈ the C-GRID as 26
$N_V$ = Number of null values

An algorithm '$A_1$' when solves the C-GRID in polynomial time '$P_t$', then '$P_t$' is taken as the reference index to refresh the key 'LTE-K' (or) the Time to Live (TTL) of the key.

Considering a time '$t_1$', which is required to compromise a single position $P1 = \{2^{32} * (nxn) * 26 * n\}$ of a grid, then the total time required to compromise the entire grid is presented in Equation (2).

$$Total\ time\ T" = \Sigma\{t_1, t_2, t_3 \ldots t_n\} \qquad (2)$$

where '$n$' is the number of positions in the C-GRID.

$$Lifetime\ of\ LTE\text{-}K = \{(Time\ taken\ to\ breach\ 2^{256}) * (iteration\ for\ the\ key\ sequence)*(Grid\ Complexity)\} \qquad (3)$$

Using Equation (3), the lifetime of the LTE-K is determined, where a polynomial algorithm '$A_1$' takes polynomial time '$P_t$' for computing the position '$P_1$' either sequentially or exponentially. So, it is theoretically impossible to find the LTE-K in the polynomial time.

The forward and backward key secrecy is maintained as it is impossible to break the key sequence besides it takes $2^{256}$ iterations to compromise a single key. So, the security level is considerably high as it uses a randomized KFF that inhibits breaking the key sequence using forward or backward keys.

$$Throughput = (No.\ of\ messages * Security\ level) / Lifetime \qquad (4)$$

$$N_k = \{(E_C - 1)*(N_c)*(N_{mc})*(E)\} \qquad (5)$$

where,

$N_k$ = Number of unique keys of C-GRID
$N_c$ = Number of columns
$N_{mc}$ = Number of mirror columns
$E$ = All possible elements in position

Equation (4) & (5) signifies the throughput and the number of unique keys generated from a C-GRID for a user respectively. Though the keys generated are considerably limited, the generated keys are random and independent enough. Thus, no relationship exists between the generated keys.

### C. Security Analysis of IMSI Protection

#### 1) Encryption and Decryption

In IPG-AKA, the cryptosystem is proposed in terms of 4-tuples as $(p, α, s, β)$, where '$p$' is a large prime number that describes which $Z^*_p$ group is used, α is an element of order n in the group $Z^*_p$, '$s$' is a random integer with '$1 \leq s \leq n-1$', and $β = α^s$. If '$p$' is primitive, then $n = p-1$. Conversely, this is not required, and '$n$' may be chosen to be smaller than '$p-1$' for increasing efficiency. The public key is $(p, α, β)$ and the private key is '$s$'.

The IMSI must be an integer between *1* and *p-1*, an element of $Z^*_p$ for encryption purpose and is split into blocks if larger than '$p-1$'. If the IMSI is a key for a symmetric cipher, then it is well known to be a member of $Z^*_p$. Further, restrictions like '*IMSI is a member of α*' can be imposed which is not recommended often. In this kind of restriction α being non-primitive, only '$n$' of the '$p-1$' members of $Z^*_p$ will be in '$α$' where no element of '$α$' can be converted to IMSI. Rising '$α$' to the power of the IMSI does not work since a discrete log is required to retrieve the original IMSI.

The encryption process requires a random integer $k \in [1, n-1]$ in addition to the public elements '$p$' and '$α$'. The encryption (E) function is shown in Equation (6).

$$E_k(IMSI) = (α^k(IMSI*β)^k)\ and\ D(u,v) = u^{-s}*v \qquad (6)$$

where, all operations are done as *mod p* in $Z^*_p$

The decryption function (D) will recover the actual IMSI from Equation (7) & (8)

$$u^{-s} = α^{-ks} = (α^s)^{-k} = β^{-k} \qquad (7)$$

$$D(E_k(IMSI)) = u^{-s}v = β^{-k} * IMSI * β^k = IMSI \qquad (8)$$

#### 2) Security

If any intruder can get the private key '$s$', it is possible to decrypt the IMSI. Since the public key includes $β = α^s$ and α, finding the private key '$s$' from the public key amounts to computing a single discrete logarithm in α. For this reason, '$n$' and '$p$' values should be very large, which is the runtime of square root discrete log algorithms and index calculus discrete log functions respectively. To break the cipher text $(u, v) = (α^k, IMSI * β^k)$, it would suffice to find $β^{-k}$ because of $IMSI = v * β^{-k}$. But $β^k$ is needed to compute inverse successfully. One can find '$k$' by computing the discrete log of $u = α^k$ base α, but this may not be significant. The complexity of cracking a cipher is equivalent to solve a Diffie-Hellman (DH) problem, given $α^k$ and $β = α^s$, determine $α^{ks} = β^k$. Solving a DH problem requires discrete log and it has been proved highly secure. Moreover, the proposed cryptosystem is too difficult to crack because of being like DH problem that is known to be highly secured.

#### 3) Efficiency

Considering a cryptosystem $(p, α, s, β)$ designed in IPG-AKA scheme, where $n = |α|$ is the order of α. The encryption process involves two exponentiations and one multiplication in $Z^*_p$. Since the exponent '$k$' is chosen between *1* and *n-1*, using a large value of '$n$' will slow down the exponentiation and in turn, slows down the encryption. A single multiplication cannot be significantly compared with the two exponentiations. Decryption involves one exponentiation, one inversion and one multiplication in $Z^*_p$. The private key '$s$' is

TABLE V
COMPARISON OF IPG-AKA WITH OTHER AKA SCHEMES

| Security Requirements | EPS AKA | Enhanced EPS-AKA | IPG AKA |
|---|---|---|---|
| Authentication | ✓ | ✓ | ✓ |
| Integrity | ✓ | ✓ | ✓ |
| Confidentiality | ✓ except for IMSI | ✓ | ✓ |
| Privacy | ✗ | ✓ | ✓ |
| Security against MitM | ✗ | ✓ | ✓ |
| Security against Replay attack | ✗ | ✓ | ✓ |
| Forward Security | ✗ | ✓ | ✓ |
| Backward Security | ✓ | ✓ | ✓ |
| Solution for single key problem | ✗ | ✗ | ✓ |
| Security against DoS attack | ✗ | ✓ | ✓ |

TABLE VI
IPG-AKA OVER EPS-AKA WITH RESPECT TO LTE-K

| Attribute | EPS-AKA | IPG-AKA |
|---|---|---|
| Key Size (bits) | 256 | 256 |
| Scalability | No | upto 768* |
| Iteration | $2^{256}$ | $2^{256*288}$ |
| Time for Key Generation (in sec) | 0.06 | 0.08 |
| Complexity Level | $O(n)$ | $O(n^3)$ |
| Key Derivation | No | Yes |
| Key Change | No | Yes |
| Key Lifetime | Permanent | Dynamic |
| Time to Compromise | Polynomial | Exponential |

*Calculated for 7x7 C-GRID

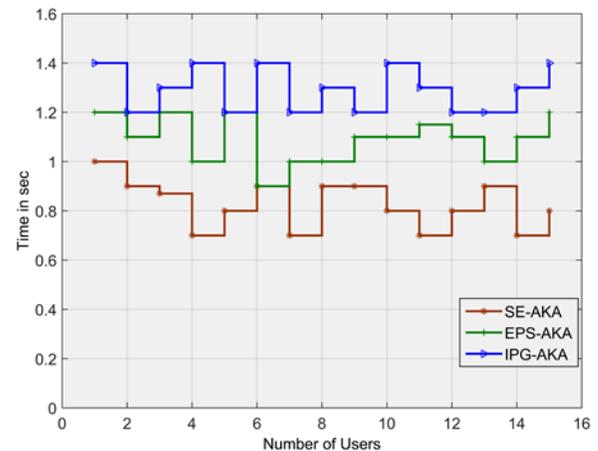

Fig. 7. Initial authentication time for AKA mechanisms

the exponent, and it is selected between *1* and *n-1*. So, choosing a smaller value for '*n*' will decrease the decryption time. Choosing a '*s*' with very lesser *1*'s in its binary representation, and by doing modular exponentiation, computation time can be significantly reduced. Multiplications involved in $Z^*_p$ are faster for a small value of '*p*', so decreasing the value of '*p*' will decrease the time required for decryption. To achieve optimal performance, developers decrease *p* where *n* = '*p-1*', or increase *p* where *n < p-1*. However, the latter choice is uncommon, and it is not preferred in practical implementation scenario as faster index calculus discrete log method depends on '*p*' and not on '*n*'.

*D. Security level comparison of the proposed scheme with existing schemes*

From Table V, unlike EPS-AKA mechanism, enhanced eEPS-AKA mechanism provides privacy and defends against MitM, replay, DoS attacks but does not solve the single key problem. Whereas, IPG-AKA protects PII and also provides an efficient solution for single key problem (via C-GRID). For the verification and efficiency of our protocol, we conducted further analysis and prove that breach time of C-GRID increases exponentially which is a guarantee of the security of our approach. However, due to space constraints, we are unable to include these results here.

## VI. EMPIRICAL PERFORMANCE ANALYSIS

Here the complexity of performing a security breach refers to the number of iterations required to find a secret key. In Table VI, we summarize the overall security aspects of EPS-AKA and the proposed IPG-AKA. Since no dependency exists between the source materials generated during IPG-AKA mechanism, forward key chaining and backward key chaining are not possible. This indicates that IPG-AKA outperforms EPS-AKA.

A statistical inference based on the observed data of initial authentication in IPG-AKA and various other authentication methods like EPS-AKA and SE-AKA are presented in Fig. 7.

It is evident that though the time required for authentication process is slightly higher, the security vulnerability preventions make it worth using. The IPG-AKA load on MME for a minimum set of users is the same initially as other protocols, but when the number of mobile users performing authentication increases, the load imposed on MME due to IPG-AKA will raise to some extent when compared to EPS-AKA and SE-AKA as shown in Fig. 8 which can be neglected with the security stand it provides. From Fig. 9, it is inferred that authentication load imposed on HSS by IPG-AKA is approximately equal to SE-AKA and EPS-AKA irrespective of the number of users and time required to perform authentication. Moreover, the LTE-K generation from the C-GRID does not impose any expensive modifications in both cost and architectural level at HSS.

As shown in the Fig. 10, it is explicit that the time taken for generating key is proportional to the size of the C-GRID. Since the difference between times required for the key generation for different grid sizes is not negligible, the size of the grid must be chosen according to the degree of the security demanded. As the grid size increases, the security also increases along with the time parameter. The bandwidth consumption for the existing EPS-AKA mechanism is compared with the proposed IPG-AKA mechanism using different set of inputs at a different mean density as shown in Fig. 11.

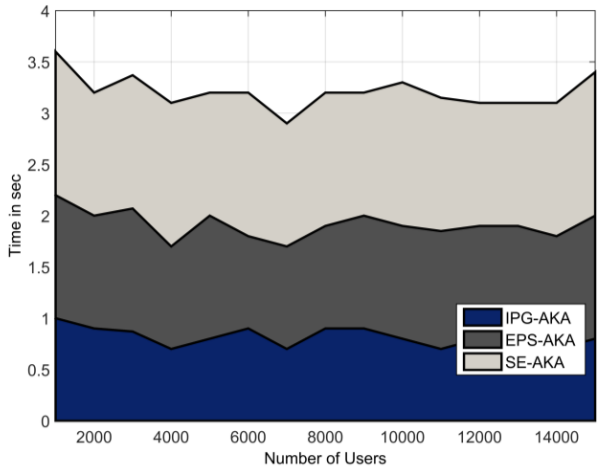

Fig. 8. Authentication load at MME

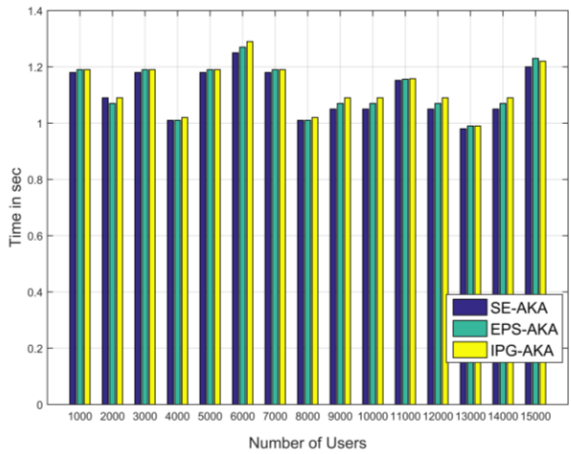

Fig. 9. Authentication load at HSS

At low mean density (say, 50/km$^2$) there is no huge difference initially but as the number of iterations increases, the bandwidth consumption of IPG-AKA gradually rises which is negligible. But at a high mean density (say, *500/km$^2$*), despite low bandwidth consumption by IPG-AKA at the start, there is a notable increase in consuming bandwidth as the iteration increases. Though the bandwidth consumption of IPG-AKA is somewhat high, it significantly protects the user identity and privacy issues making it a trustworthy mechanism.

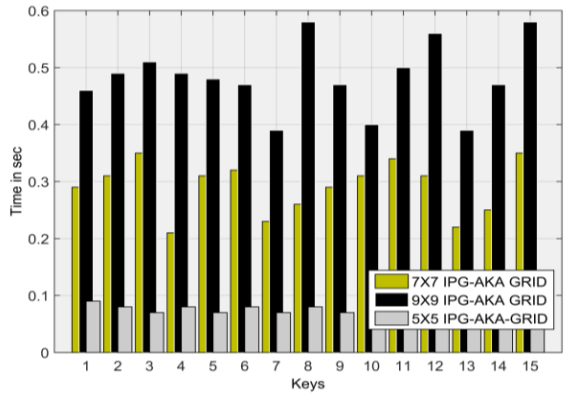

Fig. 10. Key Generation time for various size of C-GRID

TABLE VII
SECURITY ANALYSIS OF IPG-AKA WITH RESPECT TO LTE-K

| Attack Type | Intruder Knowledge | Protection |
|---|---|---|
| Chosen Cipher text | Cipher C1, C2 | Protected by Dynamic Key hierarchy |
| Known & cipher text only | | No similarity as there is no relation between the keys & Cipher text |
| Forward and Backward Secrecy | Key K1, K2, K3 | No common parameters exists, Dynamic Key sequence formation |
| Spoofing | Key K1, K2, Cipher C1, C2 and Plain Text P1 & P2 | Key freshness in KS |
| CKA | | Resolved through C-GRID complexity |
| MitM & Session Hijacking | | Dynamic key freshness in Key sequence |

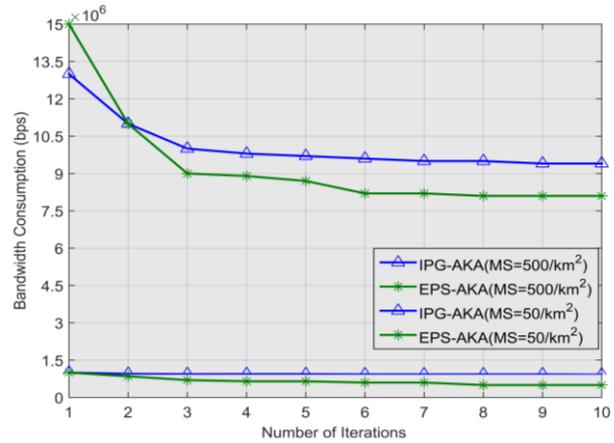

Fig. 11. Bandwidth Consumption at MME after IPG-AKA

To illustrate the security strength of our scheme, in table VII we have discussed the security provided by our scheme against the several attacks. The future computation cannot be intercepted or identified even if the primitive parameters like keying elements and other processed materials like cipher texts are known to attacker. It is because the C-GRID of the proposed IPG-AKA scheme provides a dynamic key hierarchy that adaptively changes the key sequence, which provides protection against most of the known attacks. The proposed key hierarchy is independent of the past parameters and freshness is always maintained through the dynamic source key. Forward and backward secrecy is achieved through the Key Sequence formation that dynamically alters the key fetched from the C-GRID based on the bit size. Thus, IPG-AKA framework is completely secure against active attacks like spoofing, sniffing, Compromised Key Attack (CKA), MitM attack and eavesdropping.

## VII. CONCLUSION

Owing to the broadcast nature of radio propagation channels in LTE networks, the transmission medium is susceptible to attacks such as eavesdropping, jamming, and privacy theft. So, there exists need for countermeasure development to evade these attacks by not relying on stagnant security constituents, not exposing the exploitable constructs of secure communication and making security countermeasures harder to the adversary. Thus, the proposed IPG-AKA mechanism overcomes the single key problem and privacy theft issues prevailing in the existing EPS-AKA scheme thereby providing a reliable authentication framework. The performance evaluations, computational complexity and mathematical analysis of C-GRID shows approximately 30% improvement in the security of key management. Protocol verification through AVISPA tool proves that our proposed IPG-AKA scheme is highly resistant to the LTE security vulnerabilities. In addition to that C-GRID does not require a major architectural change in the security architecture of SAE, thus it can be easily adopted and scaled. This work will significantly contribute to the PII authentication and dynamic key generation that helps to secure the transmission medium from several attacks. The future work of this article is aimed at generating the components of the C-GRID using quantum bits thereby protecting the C-GRID from quantum computers as well.

ACKNOWLEDGMENT

Rajakumar Arul gratefully acknowledges support from Anna University for Anna Centenary Research Fellowship.

Gunasekaran Raja and Rajakumar Arul, gratefully acknowledges support from NGNLab, Department of Computer Technology, Anna University, Chennai.

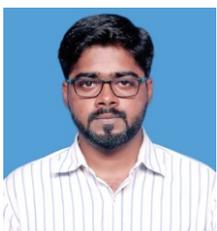

**Rajakumar Arul** (S'16 - M'17) pursued his Bachelor and Master's in Computer Science and Engineering from Anna University, Chennai. Currently, he is pursuing Doctorate of Philosophy under the Faculty of Information and Communication in NGNLab, Department of Computer Technology, Anna University - MIT Campus. He is a recipient of Anna Centenary Research Fellowship. His research interests include Security in Broadband Wireless Networks, WiMAX, LTE, Robust resource allocation schemes in Mobile Communication Networks.

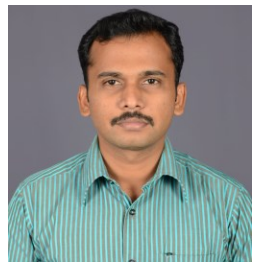

**Gunasekaran Raja** (M'08 - SM'17) is an Associate Professor in Department of Computer Technology at Anna University, Chennai and Principal Investigator of NGNLab. He received his Ph.D. in Faculty of Information and Communication Engineering from Anna University, Chennai. He was a Post-Doctoral Fellow at University of California, Davis, USA. He was a recipient of Young Engineer Award from Institution of Engineers India in 2009, Professional Achievement Award for the year 2017 from IEEE Madras Section and FastTrack grant for Young Scientist from Department of Science and Technology in 2011. Current research interest includes 5G Networks, LTE-Advanced, IoT, Wireless Security, Mobile Database, Machine Learning and Data Offloading.

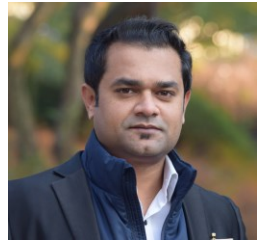

**Ali Kashif Bashir** (M'15, SM'16) is working as an Associate Professor in Faculty of Science and Technology, University of the Faroe Islands, Faroe Islands. He received his Ph.D. degree in computer science and engineering from Korea University, South Korea. In the past, he held appointments with Osaka University, Japan; Nara National College of Technology, Japan; the National Fusion Research Institute, South Korea; Southern Power Company Ltd., South Korea, and the Seoul Metropolitan Government, South Korea. He is leading several research projects and supervising/co-supervising several undergraduate and graduate (MS and PhD) students. His research interests include: cloud computing, NFV/SDN, network virtualization, network security, IoT, computer networks, RFID, sensor networks, wireless networks, and distributed computing. He has chaired several conference sessions and gave several invited and keynote talks. He is serving as the Editor-in-chief of the IEEE INTERNET TECHNOLOGY POLICY NEWSLETTER and the IEEE FUTURE DIRECTIONS NEWSLETTER.

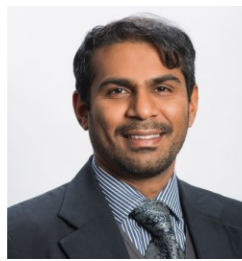

**Junaid Chaudry** is currently a cyber security faculty at College of Security and Intelligence, Embry-Riddle Aeronautical Univeristy, Prescott Arizona in USA, Junaid has over 15 years o rewarding experience in academia, industry, law- enforcement, and in corporate world in information and cyber security domain. After getting his PhD in Cyber Security from Ajou University, Junaid obtained training at Harvard Business School, University of Amsterdam, and Kaspersky Research Lab in cyber hunting and training. He is a Senior Member of IEEE, a Practicing Engineer, member of High Technology Crime Investigation Association (HTCIA), Australian Computing Society, Australian Information Security Association. His research interests include critical infrastructure protection, digital forensics, and context aware network security problems.

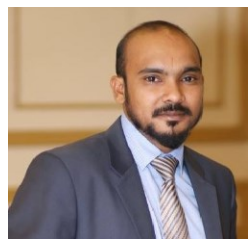

**Amjad Ali** received the Ph.D. degree in Electronics and Radio Engineering from Kyung Hee University, South Korea. Currently he is Post-Doctoral Fellow at UWB Wireless Communications Research Center in the Department of Information and Communication Engineering at Inha University, South Korea. He has published a number of research articles in refereed international journals and magazines. His main research interests include Security Issue in LTE, Internet of Things, Cognitive Radio Networks, and 5G Cellular Networks.